# Magnetic Interactions in Ball-Milled Spinel Ferrites

## G. F. Goya

Instituto de Física, Universidade de São Paulo, CP 66318, 05315-970 São Paulo, SP Brazil

Spinel $Fe_3O_4$ nanoparticles have been produced through ball milling in methyl-alcohol ($CH_3OH$), aiming to obtain samples with similar average particle sizes <d> and different interparticle interactions. Three samples having $Fe_3O_4/CH_3(OH)$ mass ratios R of 3 %, 10 % and 50 % wt. were milled for several hours until particle size reached a steady value (<d> ~ 7-10 nm). A detailed study of static and dynamic magnetic properties has been undertaken by measuring magnetization, ac susceptibility and Mössbauer data. As expected for small particles, the Verwey transition was not observed, but instead superparamagnetic (SPM) behavior was found with transition to a blocked state at $T_B$ ~ 10-20 K. Spin disorder of the resulting particles, independent of its concentration, was inferred from the decrease of saturation magnetization $M_S$ at low temperatures. For samples having 3% wt. of magnetic particles, dynamic ac susceptibility measurements show a thermally activated Arrhenius dependence of the blocking temperature with applied frequency. This behaviour is found to change as interparticle interactions begin to rule the dynamics of the system, yielding a spin-glass-like state at low temperatures for R = 50 wt.% sample.

## Introduction

The ball mill (BM) of solid/liquid mixtures has become a very common tool for making homogeneous dispersion at industrial scales, such as paint and coating production techniques. [1,2] Among the advantages of obtaining nanoparticles from BM in liquid phases are the speed of reaching a finished dispersion and the narrower distribution of particle size. However, using a carrier liquid for high-energy ball milling usually results in cemical reaction with solid particles yielding non-desired surface phases or even dissolution of the solid phase onto the liquid carrier.

Regarding the magnetic properties of such nanoparticles, the study of the mechanisms linking particle shape, size distribution and surface structure to the resulting magnetic properties is hindered by the concurrency of several mechanisms on the observed properties. These relationships are important to tailor magnetic and transport properties of nanostructured spinel ferrites $AB_2O_4$ for applications in magnetic and magnetoresistive devices. For example, the high Curie temperature ($T_C$ ~ 850 K) and nearly full spin polarization at room temperature of magnetite $Fe_3O_4$ make this material appealing for spin-devices. The bottom line is the limit imposed by superparamagnetic (SPM) relaxation in nanometric devices that curtails potential applications at room temperature. Relaxation effects are in turn strongly dependent on the magnetic dipolar interactions between particles, and thus there is an apparent need of disentangling the influence of the dipole-dipole interactions from other effects, such as surface disorder, shape anisotropy, cluster aggregation, etc. In this work, we present a study on the effects of interparticle interactions in magnetite nanoparticles dispersed in an organic liquid carrier in different concentrations, in order to study the influence of such interactions on the transition to the frozen state.

## Experimental Procedure

Three samples consisting of a dispersion of magnetite ($Fe_3O_4$) particles in a liquid carrier, with different concentrations, have been prepared by ball-milling in a planetary mill (Fristch Pulverisette 4). A mixture of magnetite powder (99.99 %, mean particle size ~0.5 μm) and methyl alcohol (between 20 and 40 ml) were introduced in a hardened steel vial and sealed in an Ar atmosphere. The samples with 50, 10 and 3 % wt. of magnetic material will be labeled F50, F10 and F3, respectively. All samples were ground for several times, extracting partial amounts after selected intervals to follow the evolution of the mean particle size <d> of the magnetite phase, stopping the experiment after reaching a plateau of <d> ~ 6-10 nm. The milling parameters and the resulting properties are summarized in Table I.

X-ray diffraction (XRD) measurements were performed using a commercial diffractometer with Cu-K$_\alpha$ radiation in the 2θ range from 10 to 80 degrees, and the morphology of the milled samples was analyzed by Transmission Electron Microscopy (TEM). Mössbauer spectroscopy (MS) measurements were performed between 20 and 294 K in dried samples. A conventional constant-acceleration spectrometer was used in transmission geometry with a [57]Co/Rh source. The recorded spectra were fitted by single-site and distribution programs, using α-Fe at 294 K to calibrate isomer shifts and velocity scale. For magnetic measurements, all samples were conditioned in closed containers before quenching the magnetite/methyl-alcohol mixture below its freezing point ( ~265 K) from room temperature. A commercial SQUID magnetometer was employed to perform static and dynamic measurements as a function field, temperature and driving frequency. Zero-field-cooled (ZFC) and field-cooled (FC) curves were taken between 5K and 250 K, to avoid the melting of the solvent in the most diluted samples. Data were obtained by first cooling the sample from room temperature in zero applied field (ZFC process) to the basal temperature, then a field was applied and the variation of magnetization was measured with increasing temperature up to T = 250

K. After the last point was measured, the sample was cooled again to the basal temperature keeping the same field (FC process); then the M *vs.* T data was measured for increasing temperatures.

## Results and Discussion

TEM images showed that all samples consisted of agglomerates of particles with average grain size $<d> \sim$ 6-10 nm (Fig. 1). The clusters of particles observed in TEM images were difficult to break with ultrasonic treatment, and thus it is not clear whether these clusters might also exist in the *as milled* dispersion. The x-ray diffractograms of the three samples along the milling series showed the same features: starting from well-defined peaks (for the initial ~0.5 μm particles), the linewidth increased steadily with increasing milling time. All patterns could be indexed with the $Fe_3O_4$ single phase (see figure 1), without any evidence of new phases formed during milling. It is worth to notice that previous results on the same particles showed the oxidation to $Fe_2O_3$ after some minutes of milling when milled in open vials. The Scherrer formula was used to estimate the $<d>$ values from each diffractogram (see Table I), without considering possible contributions of crystal stress to the observed linewidth. Estimations of $<d>$ from TEM data indicated good agreement with XRD data, although XRD data slightly underestimates the average diameter, probably due to the contribution of structural stress to the linewidth in milled samples.

The Mössbauer spectra at room temperature for all samples showed a superparamagnetic doublet with the same hyperfine parameters within experimental error (see Table I). As the temperature is decreased, a magnetic sextet develops and coexists with the SPM doublet. For T = 20 K, the spectra showed only the sextet with broadened lines (Fig. 2), which could be fitted using a hyperfine field distribution with mean values $<B_{hyp}>$ = 51.9 T, 51.3 T, and 52.2 T for samples F3, F10 and F50 respectively. The effect of thermal fluctuations for sample F3 can be still observed at T = 20 K, as a broadening on the inner side of the spectral lines, which is absent in sample F50. This difference is related to the proximity of the unblocking temperature in this sample, since the stronger dipolar interactions in F50 shift $T_B$ to higher values, as observed from magnetization data (see below).

Magnetite has a cubic spinel structure (space group Fd3m) that contains $Fe^{3+}$ ions at tetrahedral sites (*A*) and $Fe^{3+}/Fe^{2+}$ at octahedral sites (*B*), yielding ferrimagnetic order below $T_C$. [3] At low temperature (the Verwey temperature, $T_V \sim 120$ K) the systems undergoes a structural transition, and concurrently the magnetic and transport properties show a sudden change at this temperature. The ZFC/FC curves (figure 3) of the three samples show clearly that the Verwey transition is absent, and SPM behaviour is observed above the blocking temperature $T_B$ which in turn depends on the concentration.

Figure 4 displays the ac susceptibility $\chi(T,\omega)$ data of sample F3. Both real $\chi'(T)$ and imaginary $\chi''(T)$ components display a maximum at a temperature $T_m$ that depends on the driving frequency.[4] To identify the dynamic mechanisms of the freezing process, we used the empirical parameter $\Phi = \dfrac{\Delta T_m}{T_m \Delta \log_{10}(f)}$, where $\Delta T_m$ is the shift of $T_m$ within the $\Delta \log_{10}(f)$ frequency interval. This parameter provides a model-independent classification of the blocking/freezing transition. It can be seen in Table I that this value decreases strongly from for higher concentrations. The value of F3 sample is close to the $\Phi = 0.13$ expected for SPM systems. On the other side, it is known that smaller values of $\Phi$ usually result from strong interparticle interactions,[5] in agreement with the increasing concentration of samples F10 and F50 shown in Table I.

The dynamic response of an ensemble of fine particles is determined by the measuring time $\tau_m$ (or frequency) of each experimental technique. As the reversion of the magnetic moment in a single-domain particle over the anisotropy energy barrier $E_a$ is assisted by thermal phonons, the relaxation time $\tau$ exhibits an exponential dependence on temperature characterized by a Néel-Arrhenius law

$$\tau = \tau_0 \exp\left(\frac{E_a}{k_B T}\right)$$

where $\tau_0$ is in the $10^{-9}$ - $10^{-11}$ s range for SPM systems. In the absence of an external magnetic field, the energy barrier is given by $E_a = K_{eff} V$, where $K_{eff}$ is an effective anisotropy constant and V is the particle volume. It can be observed from Fig. 4 (inset) that the dependence of $\ln(f)$ *vs.* $T_B^{-1}$ is linear with a good approximation for sample F3, and accordingly the data was fitted using the Néel-Brown law yielding $E_a = 5.2 \times 10^{-21}$ J. If the average particle size from Table I is used, the resulting effective anisotropy turn out to be $K_{eff} = 46.2$ kJ/m$^3$. This value is larger than the magnetocrystalline anisotropy constant of bulk magnetite $K_1^{bulk} = 11$-$13$ kJ/m$^3$, and this is likely to originate in an extra contribution from interparticle interactions (of dipolar nature) which can also modify the effective energy barrier. Previous works on $Fe_3O_4$ particles have shown that dipolar interactions are noticeable for concentrations of ~2 % vol. of magnetic particles [6] and thus the present value of $K_{eff}$ is likely to include the effects of particle-particle interactions even for the F3 sample.

For sample F50, the temperature dependence of the cusp in $\chi_{ac}(T)$ curves is frequency-dependent (fig. 5), yielding the smaller $\Phi$ value shown in Table I. Since this low value could be originated in dipolar interactions, we have searched for spin-glass behavior, through the dynamic scaling of a.c. susceptibility by conventional critical slowing down for relaxation of the magnetic moments. When approaching the freezing temperature of a spin-glass transition from above, the characteristic relaxation time $\tau$ of individual magnetic moments will show critical slowing down, characterized by a power law $\tau \sim \xi^z$, where $\xi$ is the correlation length and $z$ is called the dynamical scaling exponent. Since $\xi$ itself has a power-law dependence on the reduced temperature $t = (T - T_C)/T_C$, where $T_C$ is the freezing temperature, in terms of $f = \tau^{-1}$ we can write

$$f \sim t^{z\nu}, \tag{I}$$

where $\nu$ is the correlation-length critical exponent. When plotted in a semi-logarithmic graph, the experimental data show the expected linear increase within the available experimental frequency range (five decades), indicating a glassy behavior. From the fit using Eq.(I) we obtained $z\nu = 21\pm2$ and $T_C = 18(2)$ K, see inset of Fig. 5. We mention here that complete determination of the critical exponents at the SG transition, performed by a full dynamic scaling of $\chi"(f,T)$, gave similar values. The obtained critical parameter product $z\nu$ is much larger than those reported for conventional 3D spin-glasses, making dubious the existence of a true transition with $T_g > 0$. Large critical exponents have been previously found in polycrystalline $(Cu_{0.75}Al_{0.25})_{1-x}Mn_x$ samples [7] and also in ball-milled $Fe_{61}Re_{30}Cr_9$ alloys [8]. Also, clear evidence of a SG phase transition through dynamic scaling analysis has been also found in concentrated Fe-C nanoparticles [9], and it was proposed that the conditions for this transition to exist are a) strong interparticle interactions and b) narrow particle-size distribution. Regarding the dynamics of spin glasses, it is known that for a broad temperature range spanning the transition temperature, each atomic moment is dynamically active, whereas for single-domain magnetic particles a variable fraction may be blocked in the experimental time window. Unless a very large interparticle distances (high dilution) is achieved in these samples, the blocked fraction of particles will act as a random field on that fraction that is unblocking (for a given temperature and time window). Our observation that for the present concentrated sample, a scaling analysis is possible to achieve is intriguing, since these ball-milled samples have a broad particle size distribution. This, in turn, should make impossible to reach the experimental situation in which the characteristic time scale associated with collective dynamics exceeds the single-particle relaxation times in the experimental time window. More work is needed to clarify this point.

In summary, we have studied the structural and magnetic properties of ball-milled magnetite particles dispersed in organic solvent, with different concentrations, showing the gradual evolution from SPM to spin-glass-like behavior. Although spin disorder is likely to occur in these samples, no definite evidence was found of enhancement of the magnetic anisotropy at the surface, nor of exchange coupling between particle surface and core. We have found that the static and dynamic properties can be understood by considering changes in the single-particle anisotropy energy $E_a$ through the effect of interparticle interactions in these concentrated systems.

## Acknowledgements

The author wishes to express his gratitude to the Brazilian agencies FAPESP and CNPq for providing partial financial support.

## FIGURE CAPTIONS

Figure 1.     X-ray diffraction pattern for sample F50. Insets show the corresponding TEM image.

Figure 2.     Mössbauer spectra of sample F3 (a) and sample F50 (b) at T = 20 K. The solid lines are the best fit to experimental data (open circles).

Figure 3.     Zero-Field Cooled (ZFC) and Field-Cooled (FC) curves for samples of different concentrations, taken with $H_{FC} = 100$ Oe.

Figure 4.     Main panel: Temperature dependence of the real component $\chi'(\omega,T)$ ac susceptibility for F3 sample, at different driven frequencies from 10 mHz to 1.5 kHz. Arrows indicate increasing frequencies. Left lower panel: Imaginary component $\chi''(T)$. Upper panel: Arrhenius plot of the relaxation time $\tau$ vs. inverse blocking temperature $T_B^{-1}$. Solid line is the best fit using Eq. (3) with $\tau_0 = 1.3 \times 10^{-14}$ s and $E_a/k_B = 378$ K.

Figure 5.    Main panel: Temperature dependence of the real component χ'(ω,T) ac susceptibility for F50 sample, at different driven frequencies from 10 mHz to 1.5 kHz. The arrow indicates the direction of increasing frequencies. Inset: log-log plot for the reduced temperature $(T - T_C)/T_C$ versus external frequency. Solid line is the best fit using eq. (I), with $T_C = 18(2)$ K and $z\nu = 21(1)$.

## Table Captions

**Table I:** Some properties of samples F3, F10 and F50: weight ratio R = 100{Fe$_3$O$_4$/CH$_3$(OH)}, ball-milling time, average particle diameters <d> (from X-ray data), blocking temperature $T_B$ (dc magnetization), relative shift of the ac susceptibility maxima Φ (see text), and hyperfine parameters at 20 and 294 K: hyperfine field ($B_{hyp}$), quadrupolar splitting (QS), Isomer shift (IS).

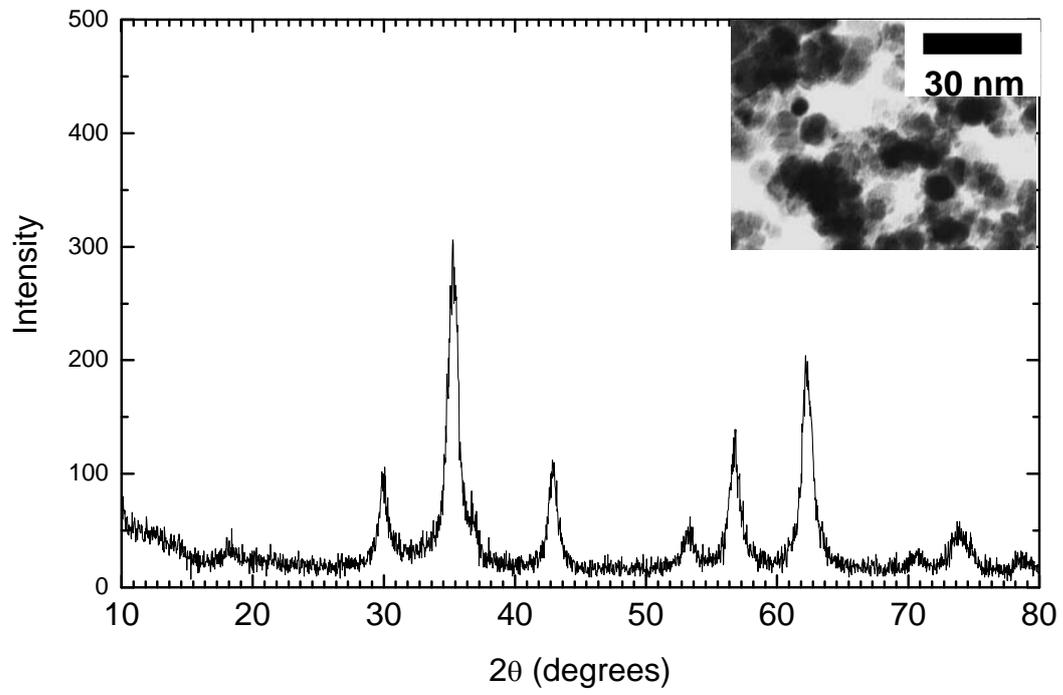

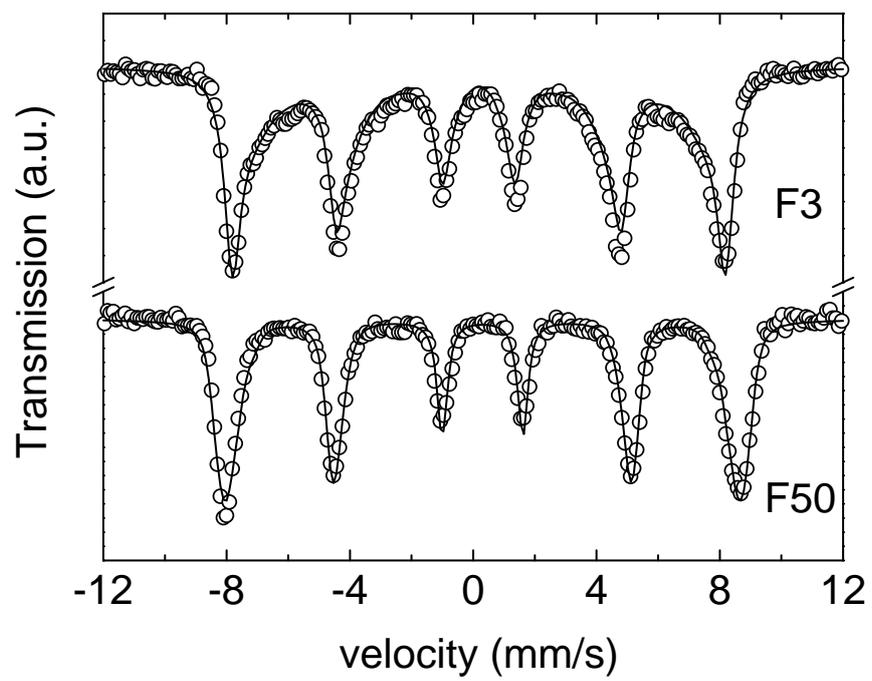

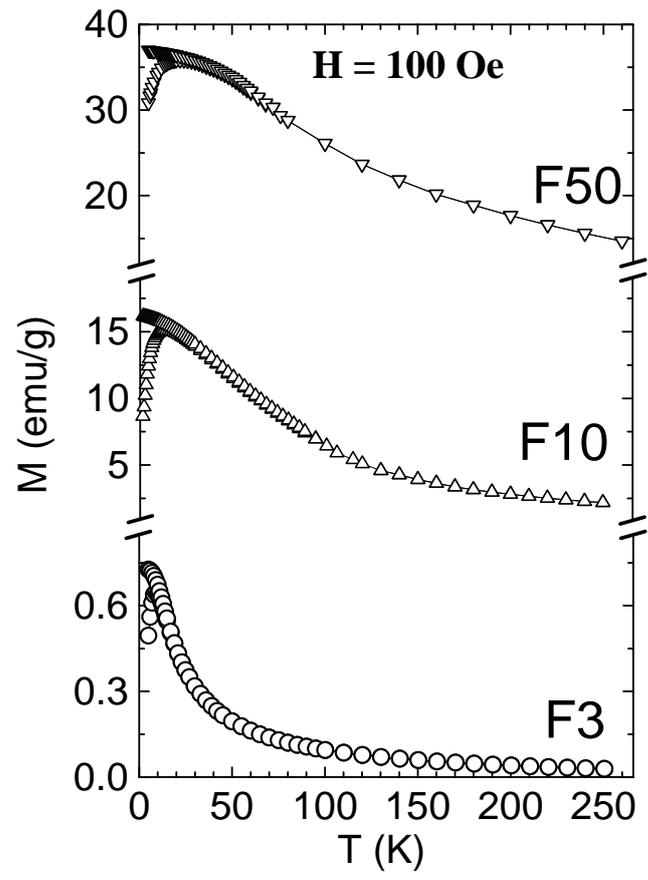

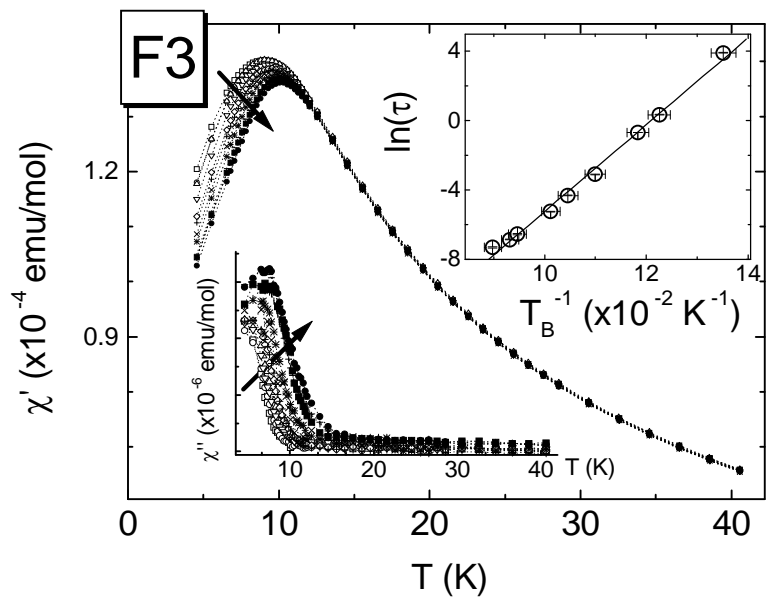

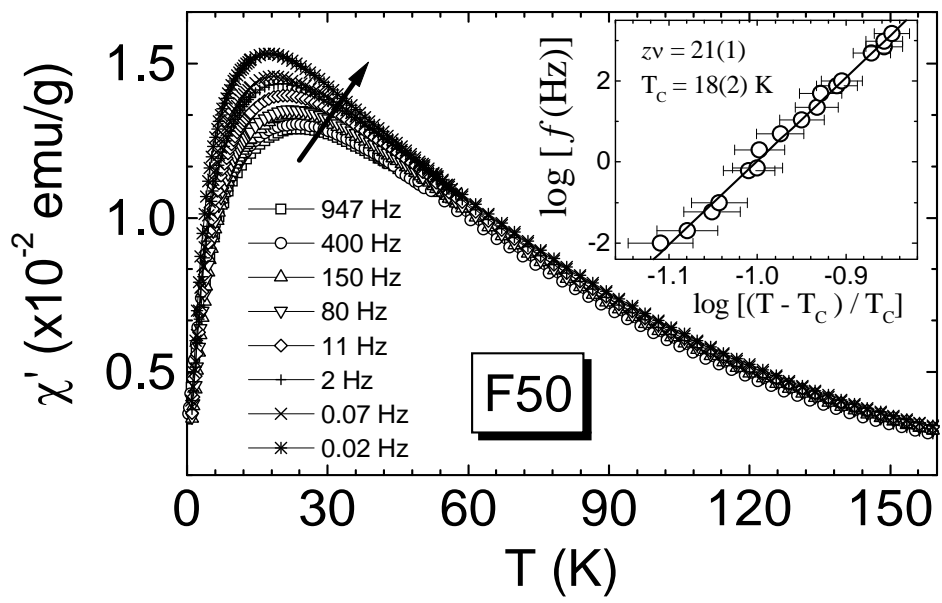